\documentclass[a4paper,11pt]{article}
\usepackage{pos}
\usepackage{graphicx}
\usepackage{cancel}
\usepackage{empheq}
\usepackage{color}
\usepackage{hyperref}

\numberwithin{equation}{section}
\usepackage{float}
\restylefloat{table}
\restylefloat{figure}
\usepackage[utf8]{inputenc}
\usepackage[font=small, labelfont=bf]{caption}
\usepackage{multirow}
\usepackage{graphicx}
\usepackage{float}
\usepackage[numbers,sort&compress]{natbib}
\usepackage{enumerate}
\usepackage{graphicx}
\usepackage{cancel}
\usepackage{empheq}
\usepackage{color}
\usepackage{colortbl}
\definecolor{green2}{cmyk}{0, 1, 0.5, 0.3}
\definecolor{green3}{cmyk}{1, 0.75, 1.0, 0.0}

\definecolor{lightgreen}{cmyk}{0.2, 0, 0.2, 0.2}
\definecolor{lightgray}{cmyk}{0.1,0.2,0,0.1}
\definecolor{lightgray2}{cmyk}{0.4,0.4,0,0.8}
\definecolor{black}{cmyk}{1.0,1.0,1.0,1.0} 

\usepackage{hyperref}
\usepackage{float}
\restylefloat{table}
\restylefloat{figure}
\usepackage{longtable}
\usepackage{lipsum} % just for dummy text- not needed for a longtable

\usepackage{comment}
\usepackage{graphicx}
\usepackage{tabu}
\usepackage{dsfont}
\usepackage{float}
\usepackage{slashed}
\usepackage{setspace}
\usepackage[labelformat=simple]{subcaption}

\usepackage{pgfplots}
\usepackage{tikz}
\usepackage{tikz-cd}
\usetikzlibrary{shapes.misc,shadows,fit,decorations.pathmorphing,positioning,trees,decorations.markings,decorations.pathreplacing,calc,shapes,patterns,arrows,positioning,arrows.meta}
\usepackage{xcolor}
\usepackage{pgfplots}
\usepackage{pgfplotstable}
\usepackage{xcolor}
\usepackage{makecell}
\usepackage{diagbox}
\usepackage{environ}
\usepackage[utf8]{inputenc}
\usepackage{multirow}
\usepackage{graphicx}
\usepackage{float}
\usepackage[numbers,sort&compress]{natbib}
\usepackage{enumerate}
\usepackage{enumitem}
\usepackage[capitalize,sort]{cleveref}
\usepackage{hhline}
\usepackage{mathtools}
\usepackage{longtable}
\usepackage{makecell}
\usepackage{tabularx}
\usepackage{cellspace}
\usepackage{alphalph}
\usepackage{lscape}
\usepackage{slashed}
\usepackage{setspace}
\usepackage{pifont}

\crefname{figure}{Figure}{Figures}
\crefname{table}{Table}{Tables}

\def\be{\begin{equation}}
\def\ee{\end{equation}}
\def\bea{\begin{eqnarray}}
\def\eea{\end{eqnarray}}
\def\ov{\overline}

\allowdisplaybreaks[2]
\numberwithin{equation}{section}

\title{On Global Embedding of Assisted Fibre Inflation}
\author*[a]{George K. Leontaris}
\author[b]{Pramod Shukla}

\affiliation[a]{Physics Department, University of Ioannina\\
	University Campus, Ioannina 45110, Greece}

\affiliation[b]{Department of Physical Sciences, Bose Institute,\\
	Unified Academic Campus, EN 80, Sector V, Bidhannagar, Kolkata 700091, India}

\emailAdd{leonta@uoi.gr}
\emailAdd{pshukla@jcbose.ac.in}

\abstract{Fibre inflation is one of the most attractive models realized in the type IIB orientifold compactification. It is embedded in the framework of L(arge) V(olume) S(cenarios) using a class of compactifying Calabi-Yau (CY) threefolds having K3-fibration. The standard single-field fibre inflation is driven by a fibre modulus which needs to travel a trans-Planckian distance of the order of ${\cal O}(5-8)$M$_p$ in the effective moduli space. The global embedding attempts using concrete CY orientifold setups have shown that K\"ahler cone conditions can generically induce some significantly tight bounds on the inflaton range, especially in the presence of a Swiss-Cheese structure via an exceptional rigid divisor in the CY threefold. Such field range bounds usually obstruct the inflationary plateau, leading to insufficient number of efolds during the inflationary dynamics.  In this context, we review our recent work about the possibility of assisting multiple fibre moduli such that the burden of traveling the required trans-Planckian distance could be shared by multiple fields, and successful inflation could be realized before hitting (or being too close to) their respective individual K\"ahler cone boundaries.}

\FullConference{Proceedings of the Corfu Summer Institute 2025 "School and Workshops on Elementary Particle Physics and Gravity" (CORFU2025)\\
27 April - 28 September, 2025\\
Corfu, Greece\\}

\tableofcontents

\begin{document}
\maketitle

\section{Introduction}
\noindent

	String theory provides a powerful and compelling framework for addressing the inscrutable conundrums of early-universe cosmology, particularly the archetype  of cosmic inflation. String theory unifies the dynamics of inflation with a consistent theory of quantum gravity, offering an ultraviolet (UV)-complete setting where the realization of inflation finds a natural theoretical  origin. Within this context, scalar moduli fields  which are ubiquitous in string compactifications, emerge as natural candidates for the inflaton, the field driving the exponential expansion. Amongst the most compelling implementations of this scenario is Fibre Inflation (FI) where the inflaton is identified with a specific K\"ahler scalar field, which we will refer to as a ``fibre'' modulus. A fascinating feature of the latter relies in its nearly flat potential, which is essential for sustained slow-roll inflation.

%String theory provides a powerful and compelling framework for addressing the profound mysteries of early-universe cosmology, particularly the paradigm of cosmic inflation. Its primary appeal lies in its capacity to unify the dynamics of inflation with a consistent theory of quantum gravity, offering a UV-complete setting where the realization of inflation finds a natural theoretical  origin. Within this setting, scalar moduli fields - ubiquitous in string compactifications- emerge as natural candidates for the inflaton, the field driving the exponential expansion.

For the predictability of any string theoretic setup designed for the phenomenological model building, moduli stabilization is a crucial step. In this regard, enormous efforts have been made since more than two decades, which have led to two most popular schemes of moduli stabilization, namely KKLT \cite{Kachru:2003aw} and LVS \cite{Balasubramanian:2005zx}. 

The common underlying idea behind both these schemes is to take a two-step approach; first step aims to stabilizing the complex structure moduli $(U^i)$ along with the axio-dilaton $(S)$ by using the leading order effects arising from the background fluxes. The K\"ahler moduli $(T_\alpha)$ which remain flat (at the leading order) due to the so-called `no-scale symmetry', are stabilized in the second step by using some non-perturbative affects. The two functions, namely a real K\"ahler potential ($K$) and a holomorphic superpotential ($W$) serve as building blocks for the purpose as they capture the low-energy dynamics of the four-dimensional effective supergravity theory. Depending on the possible corrections arising from the various sources, the K\"ahler potential and the superpotential induce useful contributions to the effective ${\cal N}=1$ scalar potential generically given as below
\begin{equation}
\label{eq:Vdef}
V_F = e^{\cal K}\left({\cal K}^{i\bar{\jmath}} D_i W \overline{D}_{\bar{\jmath}} \overline{W} - 3|W|^2 \right).
\end{equation}
The K\"ahler potential and the superpotential receive a series of corrections leading to a quite rich scalar potential structure to perform moduli stabilization and address/incorporate other issues such as inflationary cosmology and dark energy. Among such contributions, the following short list covers the ones typically used for the conventional model building: 
\begin{itemize}
\item 
The flux superpotential contributions arising from the S-dual pair of background fluxes, namely $(F_3, H_3)$ \cite{Gukov:1999ya}, are known as Gukov-Vafa-Witten (GVW) contribution. These contributions can induce leading order scalar potential piece at order ${\cal O}({\cal V}^{-2})$ in volume scaling, where ${\cal V}$ is the overall volume of the compactifying Calabi-Yau threefold. Subsequently, the GVW scalar potential can stabilize a subset of moduli, namely the complex structure moduli and the axio-dilaton.

The K\"ahler moduli remain flat at the leading order GVW scalar potential due to the so-called `no-scale symmetry' which protects their flatness. This is rooted in the fact that the volume moduli dependent piece of the tree-level K\"ahler potential, namely ${\cal K} \supset - 2 \ln{\cal V}$, satisfies an identity ${\cal K}_{\tau_\alpha}{\cal K}^{\tau_\alpha\tau_\beta}{\cal K}_{\tau_\beta} = 3$, the so-called no-scale identity, which precisely cancels the negative piece of (\ref{eq:Vdef}) at tree-level effects. Here, the four-cycle volume moduli appearing in the chiral coordinate are defined as $\tau_\alpha = \partial_{t^\alpha}{\cal V}$. We note that the overall volume of the CY threefold ${\cal V}$ depends on the two-cycle volume moduli ($t^\alpha$) as ${\cal V} = \frac16 k_{\alpha\beta\gamma}t^\alpha t^\beta t^\gamma$, where $k_{\alpha\beta\gamma}$ are triple intersection numbers.

\item 
The no-scale symmetry is broken by the so-called BBHL corrections \cite{Becker:2002nn} which are induced through higher derivative effects appearing at $(\alpha^\prime)^3$ order. The effects of such a correction changes the volume dependent K\"ahler potential piece with a shift in the overall volume of the CY threefold, i.e. ${\cal K} \supset - 2 \ln\left({\cal V}+\xi/2\right)$, where $\xi$ encodes information about the Calabi-Yau threefold. This subsequently induces an scalar potential piece at subleading order which turns out to be ${\cal O}({\cal V}^{-3})$ in the inverse volume scaling.

\item 
The BBHL correction although breaks the no-scale structure, one still needs contributions of additional types in order to achieve a viable volume moduli stabilization. The two schemes, namely KKLT and LVS uses the non-perturbative contributions to the holomorphic superpotential \cite{Witten:1996bn}. These contributions are induced via E3-instantons or gaugino condensation effects through stacks of D7-branes, by wrapping suitable rigid divisors of the compactifying Calabi-Yau threefold. The KKLT scheme of moduli stabilization \cite{Kachru:2003aw} uses such corrections to get a supersymmetric minimization of all the moduli in AdS while LVS uses a combination of competing BBHL and non-perturbative superpotential effects to achieve an exponentially large VEV for the overall volume modulus \cite{Balasubramanian:2005zx}.

Apart from the non-perturbative superpotential effects proposed in \cite{Witten:1996bn,Green:1997di,Blumenhagen:2009qh, Blumenhagen:2008zz}, there can be other sources of non-perturbative effects such as poly-instanton corrections \cite{Blumenhagen:2012kz}, and also those which arise after `rigidifying' a non-rigid divisor using magnetic fluxes as proposed/studied in series of works \cite{Bianchi:2011qh,Bianchi:2012pn,Louis:2012nb}.

\item 
After minimizing the overall volume ${\cal V}$ along with the complex-structure moduli and the axio-dilaton in a given scheme, say LVS, there can be several moduli which still remain flat and can serve for the purpose of realizing inflation. These so-called ``LVS flat directions" are potential inflaton candidates. Depending on the specific model dependent features, one can have several classes of inflationary models such as Blow-up inflation or K\"ahler moduli inflation \cite{Conlon:2005ki,Blanco-Pillado:2009dmu,Cicoli:2017shd}, Fibre Inflation \cite{Cicoli:2008gp,Cicoli:2016chb,Cicoli:2016xae,Cicoli:2017axo, Cicoli:2024bxw}, Poly-intanton inflation \cite{Cicoli:2011ct,Blumenhagen:2012ue,Gao:2013hn,Gao:2014fva}, and combinations thereof~\cite{Bansal:2024uzr}. 

\item 
For stabilizing the LVS flat directions, the string loop corrections are extremely useful. There are broadly two classes of string loop corrections; one follows from the prescription of \cite{Berg:2004ek,vonGersdorff:2005bf, Berg:2005ja, Berg:2005yu, Cicoli:2007xp, Gao:2022uop} where presence non-intersecting D7/O7 configurations typically leads to ``KK-type" while the intersecting stacks of D7/O7 with non-contractible intersection loci lead to the so-called ``Winding-type" string loop corrections.

Another class of perturbative string-loop effects turns out to be the so-called ``log-loop effect" \cite{Antoniadis:2018hqy,Antoniadis:2019rkh,Antoniadis:2020ryh}, which has attracted attention in recent years. The reason being the fact that using a combination of BBHL and log-loop corrections one can minimize the overall volume of the CY threefold to exponentially large values in the weak coupling regime. Due to heavily being similar to the conventional LVS, this scheme of moduli stabilization is called as ``perturbative LVS" \cite{Leontaris:2022rzj} which does not need a Swiss-Cheese structure for the CY threefold.

\item
Finally, there is another higher derivative correction to the scalar potential which appears at F$^4$-order, where F denotes the F-terms \cite{Ciupke:2015msa}. Such a correction falls beyond the two derivative description in terms of the K\"ahler potential and the superpotential captured from the expression (\ref{eq:Vdef}). However, these corrections also appear at $(\alpha^\prime)^3$ order in the perturbation series, like the BBHL piece.

\end{itemize}

Among the most compelling inflationary realizations in LVS is the Fibre Inflation, a rigorously derived model where the inflaton is identified as a specific Kähler modulus, known as a ``fibre'' modulus. Crucially, at leading order its flat potential, which is essential for sustained slow-roll inflation, is not reliant on fine-tuned supersymmetry but is generated and stabilized through well-defined robust perturbative corrections within the string theory landscape. It eventually turns out after stabilizing the volume $\mathcal{V}$ and the axionic partners at the leading order (perturbative) LVS, the effective single-field potential for the canonical fibre inflaton $\phi$ can be given as
\begin{equation}
V(\phi) = V_0 \left( 3 - 4 e^{-\phi/\sqrt{3}} + e^{-4\phi/\sqrt{3}} + \dots \right),
\end{equation}
which exhibits a plateau region suitable for slow-roll inflation. Here, the overall coefficient $V_0$ is a model dependent parameter which is fixed while matching the scalar perturbation amplitude at the horizon exit. The slow-roll parameters,
\begin{equation}
\epsilon = \frac{M_{\text{Pl}}^2}{2} \left( \frac{V'}{V} \right)^2, \quad \eta = M_{\text{Pl}}^2 \frac{V''}{V},
\end{equation}
remain small over a super-Planckian field range $\Delta \phi > M_{\text{Pl}}$, yielding sufficient efolds during inflation and a scalar spectral index $n_s$ and tensor-to-scalar ratio $r$ consistent with current observational bounds.

This introduction sets the stage for exploring how Fibre Inflation embodies the successful merger of inflationary cosmology with the fundamental principles of string theory, providing a concrete model where the inflaton potential is derived from first principles and connects directly to the geometry of extra dimensions. Our main aim is to review the recent work \cite{Leontaris:2025hly} on multi-field fibre inflation in perturbative LVS where we show that the burden of creating sufficient efolds can be shared by multiple fibre moduli. This result is of great interest, in particular, in the light of recent challenges faced in the conventional single-field fibre inflation models of the standard Swiss-Cheese LVS \cite{Cicoli:2018tcq,Bera:2024ihl}.

\section{Moduli Stabilization in LVS}
\noindent
In type IIB superstring theory compactified on a Calabi-Yau threefold, the four-dimensional effective theory contains a multitude of scalar fields, known as moduli, which parameterize the geometry of the extra dimensions. Stabilizing these moduli—i.e., generating a potential with finite minima for their vacuum expectation values—is essential for obtaining a realistic low-energy physics and a stable cosmological background. This process unfolds in several distinct steps, beginning with the flux-induced superpotential.

\subsection{The Flux-Induced Superpotential}
\noindent
The primary tool for stabilizing complex structure moduli and the axio-dilaton is the Gukov-Vafa-Witten (GVW) superpotential, generated by turning on three-form fluxes in the internal manifold. At tree-level, this superpotential is given by
\begin{equation}
    {W}_{\rm flux} = \int_{{\rm CY_3}} \mathbf{G}_3 \wedge \Omega(U^i),
    \label{eq:W0}
\end{equation}
which depends crucially on three ingredients:
\begin{itemize}
    \item The complex three-form flux $\mathbf{G}_3 = F_3 - S H_3$, a combination of the Ramond-Ramond ($F_3$) and Neveu-Schwarz ($H_3$) field strengths.
    \item The axio-dilaton modulus $S = C_0 + i e^{-\phi}$, which combines the Ramond-Ramond zero-form $C_0$ and the dilaton $\phi$.
    \item The holomorphic $(3,0)$-form $\Omega$, which depends on the complex structure moduli $U^i$ that describe the shape of the Calabi-Yau manifold.
\end{itemize}
This superpotential, ${W}_{\rm flux}$, is a holomorphic function of $S$ and $U^i$, but notably \textit{does not depend} on the K\"ahler moduli $T_\alpha$ at this classical level.

\subsection{Stabilizing Complex Structure Moduli and the Axio-Dilaton}
\noindent
The fluxes induce a potential for $S$ and $U^i$ via the standard $\mathcal{N}=1$ supergravity F-term scalar potential. Supersymmetric minima are found by solving the F-flatness conditions $D_I {W}_{\rm flux} = 0$, where $D_I = \partial_I + (\partial_I \mathcal{K})$ is the K\"ahler covariant derivative. This leads to two key conditions:
\begin{enumerate}
    \item \textbf{Axio-dilaton stabilization:}
    \begin{equation}
        D_S {W}_{\rm flux} = 0 \quad \implies \quad \int_{{\rm CY_3}} \overline{G}_3 \wedge {\Omega} = 0.
        \label{eq:DSW}
    \end{equation}
%This condition essentially forces $G_3$ to be a primitive $(2,1)$-form.
    \item \textbf{Complex structure moduli stabilization:}
    \begin{equation}
        D_{U^i} {W}_{\rm flux} = 0 \quad \implies \quad \int_{{\rm CY_3}} G_3 \wedge \chi_i = 0,
        \label{eq:DUW}
    \end{equation}
where $\chi_i$ are the basis forms for $(2,1)$ cohomology. In addition, unborken supersymmetry requires $D_{T_\alpha} W_{\rm flux} = 0$, which is equivalent to $W_{\rm flux}=0$. These three conditions essentially force $G_3$ to be a primitive $(2,1)$-form. Subsequently, all complex structure moduli $U^i$ and the axio-dilaton $S$ are fixed at a high mass scale, typically near the Kaluza-Klein scale.
\end{enumerate}
%Provided the flux tadpole constraints are satisfied (see, e.g., \cite{Bena:2019...}), $S$ and $U^i$ are completely stabilized. 
However, a major challenge remains: the K\"ahler moduli, which govern the volumes of cycles within the Calabi-Yau, are \textbf{not fixed} by this tree-level potential. This is because they do not appear in ${W}_{\rm flux}$.
Despite their presence in the tree-level K\"ahler potential $\mathcal{K}_0$, the resulting F-term scalar potential vanishes identically due to the famous \textit{no-scale structure} of type IIB compactifications:
\begin{equation}
    V = e^{\mathcal{K}_0} \left( \sum_{I,J} \left(D_I {W}_{\rm flux}\right) \, \mathcal{K}_0^{I\bar{J}} \, \left(\overline{D}_{\bar{J}} \overline{{W}}_{\rm flux}\right) - 3 |{W}_{\rm flux}|^2 \right) \equiv 0,
    \label{eq:no-scale}
\end{equation}
when conditions $D_S \,{W}_{\rm flux} = D_{U^i} {W}_{\rm flux} = 0$ are imposed. This cancelation highlights the need for additional effects to lift the flat directions associated with the K\"ahler moduli.

\subsection{K\"ahler Moduli Stabilization}
\noindent
To generate a potential for the K\"ahler moduli $T_\alpha$, one must go beyond the tree-level approximation. The tree-level K\"ahler potential $\mathcal{K}_0$ is corrected by both $\alpha'$ and quantum loop effects. The leading $\alpha'^3$ correction, proportional to the Euler characteristic $\chi$ of the Calabi-Yau, was computed in \cite{Becker:2002nn}. Furthermore, perturbative logarithmic corrections of the form $\ln(\mathcal{V})$ have also been identified \cite{Antoniadis:2018hqy,Antoniadis:2019rkh,Antoniadis:2020ryh}. Incorporating these, the corrected K\"ahler potential and the superpotential take the following form:
\be
\label{eq:K-W-gen}
{\mathcal K} = -\ln\left[-i\int \Omega\wedge\bar{\Omega}\right]-\ln\left[-\,i\,(S-\bar{S})\right]-2\ln{\cal Y}, \qquad W= W_{\rm flux} + W_{\rm np}\,,
\ee
where $\Omega$ denotes the nowhere vanishing holomorphic 3-form of the compactifying Calabi-Yau threefold which depends on the complex-structure moduli, while ${\cal Y}$ encodes several contributions. For the superpotential, $W_{\rm flux}$ is induced by usual S-dual pair of the 3-form fluxes $(F_3, H_3)$ \cite{Gukov:1999ya} which depends on the $\{S, U^i\}$ moduli while the non-perturbative corrections $W_{\rm np}$ \cite{Witten:1996bn} can have $T_\alpha$ dependence to break the  no-scale symmetry needed to facilitate the volume moduli stabilization. Another alternative route is to consider some more rich structure in the K\"ahler potential ${\cal K}$, say via ${\cal Y}$ which can receive a variety of (non-)perturbative corrections. Three primary mechanisms are employed, often in conjunction:
\begin{enumerate}
\item \textbf{Perturbative $\alpha^\prime$-corrections to the K\"ahler potential:} The leading order $\alpha^\prime$-corrections appear as shifted volume in the tree-level K\"ahler potential \cite{Becker:2002nn} with ${\cal Y}$ given as, 
\bea
\label{eq:YwithBBHL}
& & {\cal Y} = {\cal V} +  \frac{\xi}{2} \, e^{-\frac{3}{2} \phi}\,,
\eea
where $\xi=-\frac{\chi(X)\,\zeta(3)}{2\,(2\pi)^3}$, and $\chi(X)$ denotes the CY Euler characteristic. In fact these corrections appear at order $\alpha'^3$ and are also referred as BBHL corections.

\item \textbf{Non-perturbative contributions to the superpotential:}
Non-perturbative effects, such as Euclidean D3-brane instantons or gaugino condensation on D7-branes, generate an explicit dependence on the K\"ahler moduli $T_i = \tau_i + i \theta_i$ (where $\tau_i$ are volumes and $\theta_i$ are axions). The superpotential becomes
\begin{equation}
{W} = {W}_{0} + \sum_k \mathcal{A}_k e^{-a_k T_k},
\label{eq:Wnp}
\end{equation}
where ${W}_0$ is the flux-stabilized constant $\langle W_{\rm flux}\rangle$ from Eq.~\eqref{eq:W0}, $\mathcal{A}_k$ are prefactors, and $a_k$ are constants (e.g., $2\pi$ for E3-instantons).

The interplay between the corrected K\"ahler potential \eqref{eq:K-W-gen} with ${\cal Y}$ given in (\ref{eq:YwithBBHL}), and the non-perturbative superpotential \eqref{eq:Wnp} breaks the no-scale structure and generates a scalar potential for the K\"ahler moduli. This potential can stabilize all geometric moduli in a metastable de Sitter or Minkowski vacuum—a foundational achievement known as the \textit{KKLT} or \textit{Large Volume Scenario} (LVS), depending on the dominant balancing terms.

\item \textbf{Perturbative string loop corrections: An alternative to non-perturbative effects:}
The presence of non-perturbative superpotential term is essential part of both the moduli stabilization schemes, namely KKLT and LVS. In a given Calabi-Yau orientifold setup, such a correction is not generically present and needs some rigid (exceptional) or rigidified divisors wrapping the E3-instantons or suitable D7-brane stacks for the gaugino condensation. This process, subsequently, ends up in discarding a huge number of Calabi-Yau threefolds for phenomenological model building. In this regard, the so-called `log-loop' corrections have attracted significant interests in the recent few years. Incorporating these effect, the corrected ${\cal Y}$ can be given as follows
\bea
\label{eq:YwithLogLoops}
& & {\cal Y} = {\cal V} +  \frac{\xi}{2} \, e^{-\frac{3}{2} \phi} + e^{\frac{1}{2} \phi}\, \left(\sigma + \eta \, \ln{\cal V}\right)\,,
\eea
where first two terms with tree-level ${\cal V}$ and the BBHL shift with $\xi$ \cite{Becker:2002nn} are zeroth order in the string-loop expansion while the last piece is the one-loop correction. The explicit presence of dilaton in the BBHL piece is merely a reflection of the expression being written in the Einstein frame. The other model dependent parameters are given as
\bea
\label{eq:def-xi-eta}
& & \hskip-1cm  \xi = - \frac{\chi({\rm CY})\, \zeta[3]}{2(2\pi)^3}~, \,\, \sigma  = - \frac{\chi({\rm CY})\, \zeta[2]}{2(2\pi)^3} \sigma_0, \,\, \eta =  \frac{\chi({\rm CY})\, \zeta[2]}{2(2\pi)^3} \eta_0, \,\, \frac{\xi}{\eta} = -\frac{\zeta[3]}{\zeta[2]} 
\eea
We note that, the parameters $\xi$, $\sigma$ and $\eta$ do not depend on the string coupling $g_s$ following from the $SL(2, {\mathbb Z})$ arguments, and should be fixed after first step of moduli stabilization \cite{Leontaris:2022rzj}. However, $\sigma$ and $\eta$ parameters generically may have a dependence on the complex structure moduli, and for that purpose we have introduced two new parameters, namely $\sigma_0$ and $\eta_0$, while still keeping the $SL(2,\mathbb Z)$ motivated factors appearing with the Riemann $\zeta$-functions and the Euler characteristic of the CY threefold.

\end{enumerate}
%

%%%%%%%%%%%%%%%%%%%%%%%%%%%%%%%%%%%%%%%%%%%%%%%%%%%%%%%%%%%%%%%%%%%%%%%%%%%%
%%%%%%%%%%%%%%%%%%%%%%%%%%%%%%%%%%%%%%%%%%%%%%%%%%%%%%%%%%%%%%%%%%%%%%%%%%%%

\section{Perturbative LVS: A Novel Approach to Moduli Stabilization}

%%%%%%%%%%%%%%%%%%%%%%%%%%%%%%%%%%%%%%%%%%%%%%%%%%%%%%%%%%%%%%%%%%%%%%%%%%%%
%%%%%%%%%%%%%%%%%%%%%%%%%%%%%%%%%%%%%%%%%%%%%%%%%%%%%%%%%%%%%%%%%%%%%%%%%%%%
\noindent
String inflation models face the significant challenge of consistently stabilizing all moduli fields while generating a flat potential suitable for slow-roll inflation. Fibre inflation, embedded within the framework of perturbative Large Volume Scenario (pLVS), provides a compelling solution that operates without relying on non-perturbative effects like instantons or gaugino condensation. This is particularly advantageous in Calabi-Yau manifolds that lack rigid divisors, which are typically required for such non-perturbative contributions.

The simplest realization of fibre inflation in pLVS is based on a K3-fibred Calabi-Yau orientifold whose volume form mimics that of a toroidal compactification \cite{Gao:2013pra,Leontaris:2022rzj}:
\begin{equation}
\mathcal{V} \propto \sqrt{\tau_1 \tau_2 \tau_3},
\label{eq:toroidal_volume}
\end{equation}
where $\tau_\alpha$ are the volumes of four-cycles (divisors) in the Calabi-Yau threefold. The absence of rigid divisors in this specific geometry prevents the use of standard moduli stabilization schemes like KKLT or LVS, necessitating the perturbative approach of pLVS.

%%%%%%%%%%%%%%%%%%%%%%%%%%%%%%%%%%%%%%%%%%%%%%%%%%%%%%%%%%%%%%%%%%%%%%%%%%%%%
\subsection{Global Model Specification}
%%%%%%%%%%%%%%%%%%%%%%%%%%%%%%%%%%%%%%%%%%%%%%%%%%%%%%%%%%%%%%%%%%%%%%%%%%%%%
\noindent
The Calabi-Yau threefold is defined as a hypersurface in a 4-dimensional toric variety, described by the following toric GLSM data:
\begin{center}
\begin{tabular}{|c|ccccccc|}
\hline
\cellcolor[gray]{0.9}Hyp &\cellcolor[gray]{0.9} $x_1$  &\cellcolor[gray]{0.9} $x_2$  &\cellcolor[gray]{0.9} $x_3$  &\cellcolor[gray]{0.9} $x_4$  &\cellcolor[gray]{0.9} $x_5$ & \cellcolor[gray]{0.9}$x_6$  &\cellcolor[gray]{0.9} $x_7$       \\
\hline
\cellcolor[gray]{0.9}4 & 0  & 0 & 1 & 1 & 0 & 0  & 2   \\
\cellcolor[gray]{0.9}4 & 0  & 1 & 0 & 0 & 1 & 0  & 2   \\
\cellcolor[gray]{0.9}4 & 1  & 0 & 0 & 0 & 0 & 1  & 2   \\
\hline
& K3  & K3 & K3 &  K3 & K3 & K3  &  SD  \\
\hline
\end{tabular}
\end{center}
The columns $x_1$ to $x_7$ represent seven toric divisors, with the first six being K3 surfaces and the seventh a Special Deformation (SD) divisor. The "Hyp" rows define the degree of the hypersurface equation $P(x_1, \dots, x_7) = 0$ that specifies the Calabi-Yau manifold. The Stanley-Reisner ideal, which lists coordinate combinations that cannot vanish simultaneously, is:
\begin{equation}
\mathrm{SR} = \{x_1 x_6, \, x_2 x_5, \, x_3 x_4 x_7\}.
\end{equation}
This encodes intersection properties: $D_1$ and $D_6$ do not intersect, $D_2$ and $D_5$ do not intersect, and $D_3$, $D_4$, and $D_7$ have no common intersection point.

%%%%%%%%%%%%%%%%%%%%%%%%%%%%%%%%%%%%%%%%%%%%%%%%%%%%%%%%%%%%%%%%%%%%%%%%%%%%%
\subsubsection*{Hodge Numbers and Topology}
\noindent
The Calabi-Yau has Hodge numbers:
\begin{itemize}
\item $h^{1,1} = 3$: Three K\"ahler moduli, corresponding to independent deformations of the K\"ahler form. For favorable CY geometries, this count follows as: $h^{1,1} = \#\text{Toric Divisors} - \#\text{Linear Relations} - \#\text{CY Hypersurface Constraints} = 7 - 3 - 1 = 3$.
\item $h^{2,1} = 115$: Complex structure moduli controlling the shape of the defining polynomial.
\item Euler number: $\chi = 2(h^{1,1} - h^{2,1}) = -224$.
\end{itemize}
We find that six of the toric divisors are $K3$ surfaces, $D_i = K3$ for $i \in \{1, 2, .., 6\}$ with a single deformation, while the $D_7$-divisor has more deformations and happens to be of the so-called ``$SD$ type" in the conventional classification \cite{Gao:2013pra, Shukla:2022dhz}. The various divisor topologies are characterized by their Hodge diamonds:
\begin{equation}
K3 \equiv \begin{tabular}{ccccc}
& & 1 & & \\
& 0 & & 0 & \\
1 & & 20 & & 1 \\
& 0 & & 0 & \\
& & 1 & & \\
\end{tabular}, \qquad \qquad
SD \equiv \begin{tabular}{ccccc}
& & 1 & & \\
& 0 & & 0 & \\
27 & & 184 & & 27 \\
& 0 & & 0 & \\
& & 1 & & \\
\end{tabular}.
\end{equation}

%%%%%%%%%%%%%%%%%%%%%%%%%%%%%%%%%%%%%%%%%%%%%%%%%%%%%%%%%%%%%%%%%%%%%%%%%%%
\subsubsection*{Intersection Numbers and Volume Form}
\noindent
Choosing the smooth divisor basis $\{D_1, D_2, D_3\}$ and their duals $\{\hat{D}_1, \hat{D}_2, \hat{D}_3\}$, the intersection polynomial for this CY threefold is simply given as:
\begin{equation}
I_3 = 2\, \hat{D}_1\, \hat{D}_2\, \hat{D}_3,
\end{equation}
with the only non-zero intersection number $k_{123} = 2$. Considering the K\"ahler form as:
\begin{equation}
J = t^1\hat{D}_1 + t^2\hat{D}_2 + t^3\hat{D}_3,
\end{equation}
the Calabi-Yau volume becomes:
\begin{equation}
\mathcal{V} = \frac{1}{3!} \int J \wedge J \wedge J = \frac{1}{6} \, \kappa_{ijk} t^i t^j t^k = 2t^1t^2t^3 = \frac{\sqrt{\tau_1\tau_2\tau_3}}{\sqrt{2}},
\end{equation}
with K\"ahler cone conditions being given as: $t^1 > 0, \, t^2 > 0, \, t^3 > 0$.

%%%%%%%%%%%%%%%%%%%%%%%%%%%%%%%%%%%%%%%%%%%%%%%%%%%%%%%%%%%%%%%%%%%%%%%%%%%
\subsubsection*{Choice of Involution and Brane Setting}
\noindent
Having the largest GLSM charge among the seven toric divisors, we consider a holomorphic involution which flips the $x_7$ coordinate, i.e., $\sigma_7: x_7 \to - x_7$. This turns out to give only one fixed point set with $\{O7 = D_7\}$ and there are no $O3$-planes present. Now considering a brane configuration of three stacks of $D7$-branes wrapping each of the three divisors $\{D_1, D_2, D_3\}$ leads to the following D7-brane tadpole cancellation,
\bea
& & 8\, [O_7] = 8 \left([D_1] + [D_1^\prime] \right) + 8 \left([D_2] + [D_2^\prime] \right) + 8 \left([D_3] + [D_3^\prime] \right)\,,
\eea
along with the $D3$ tadpole cancellation condition resulting in the following constraints:
\bea
\label{eq:ND3-tadpole}
& & \hskip-1cm N_{\rm D3} + \frac{N_{\rm flux}}{2} + N_{\rm gauge} = \frac{N_{\rm O3}}{4} + \frac{\chi({\rm O7})}{12} + \sum_a\, \frac{N_a \left(\chi(D_a) + \chi(D_a^\prime) \right) }{48} \\
& & \hskip2.4cm = 0 + \frac{240}{12} + 8 + 8 + 8 = 44\,,\nonumber
\eea
which results in some fair amount of flexibility in the required tuning of the background (and gauge) fluxes contributing through $N_{\rm flux}$ and $N_{\rm gauge}$ appearing in the left side of the equation (\ref{eq:ND3-tadpole}). Moreover, it also helps in choosing relatively larger values for the flux superpotential VEV, namely $W_0$, while building the numerical models. In fact, $W_0$ parameter is intertwined with the D3 tadpole charge $Q_{\rm D3}$ as: $(2\pi g_s \, |W_0|^2) < Q_{\rm D_3} = 88$ \cite{Denef:2004ze,Cicoli:2024bxw}, where $Q_{\rm D_3}$ is twice the right-hand-side (rhs) of (\ref{eq:ND3-tadpole}), for example see \cite{Blumenhagen:2008zz}. 

%%%%%%%%%%%%%%%%%%%%%%%%%%%%%%%%%%%%%%%%%%%%%%%%%%%%%%%%%%%%%%%%%%%%%%%%%%%%%
%%%%%%%%%%%%%%%%%%%%%%%%%%%%%%%%%%%%%%%%%%%%%%%%%%%%%%%%%%%%%%%%%%%%%%%%%%%%%

\subsection{Induced Contributions to the Scalar Potential}
%%%%%%%%%%%%%%%%%%%%%%%%%%%%%%%%%%%%%%%%%%%%%%%%%%%%%%%%%%%%%%%%%%%%%%%%%%%%%
%%%%%%%%%%%%%%%%%%%%%%%%%%%%%%%%%%%%%%%%%%%%%%%%%%%%%%%%%%%%%%%%%%%%%%%%%%%%%
\noindent
The specific underlying geometry following from the choice of holomorphic involution, and the subsequent Brane setting induces several corrections to the effective potential:

\begin{itemize}

\item \textbf{Background flux contributions:} The standard S-dual fluxes $(F_3, H_3)$ induce the leading order contributions to the four-dimensional effective scalar potential via the holomorphic flux superpotential $W_{\rm flux}$. These contributions facilitate the fixing of complex structure moduli and the axio-dilaton in their respective supersymmetric minima. 

\item \textbf{Leading order `No-scale' breaking corrections:} The K\"ahler moduli remain flat in the presence of background fluxes. However, there are several ways to break the no-scale symmetry such as higher derivative BBHL corrections \cite{Becker:2002nn}, and other perturbative \cite{Berg:2004ek,vonGersdorff:2005bf, Berg:2005ja, Berg:2005yu,Cicoli:2007xp, Gao:2022uop,Antoniadis:2018hqy,Antoniadis:2019rkh,Antoniadis:2020ryh} and non-perturbative effects \cite{Witten:1996bn}. For our current interest, we consider the `log-loop' effects of perturbative type, which leads to the following leading-order no-scale breaking term,
\bea
\label{eq:VpLVSsimp}
& & V_{\rm pLVS} \simeq \frac{3\,\kappa\, |W_0|^2}{4 {\cal V}^3}\left(\hat{\xi} + 2\hat{\eta}\ln\mathcal{V} - 8\hat{\eta} + 2\hat{\sigma}\right),
\eea
where using (\ref{eq:def-xi-eta}), the hatted quantities $\hat\xi$, $\hat\eta$ and $\hat\sigma$ are defined as 
\bea
\label{eq:def-xi-eta1}
& & \hskip-1cm \hat\xi = \frac{\xi}{g_s^{3/2}}~, \quad \hat\sigma = g_s^{1/2}\, \sigma~,\quad \hat\eta = g_s^{1/2}\, \eta~, \quad \frac{\hat\xi}{\hat\eta} = -\frac{\zeta[3]}{\zeta[2]\,g_s^2\, \eta_0}, \quad \frac{\hat\sigma}{\hat\eta} = -\frac{\sigma_0}{\eta_0}~.
\eea
We recall again that the absence of any rigid divisor makes the non-perturbative effects generically absent in this setup, and there are no such corrections to break the no-scale symmetry. 

\item \textbf{Winding-type corrections:} Following the recipe of \cite{Berg:2004ek,vonGersdorff:2005bf, Berg:2005ja, Berg:2005yu}, D7-brane stacks and $O7$-plane stacks wrapping suitable four-cycles leading to intersection loci as some non-contractible curves (such as two-tori $\mathbb{T}^2$ or higher genus curves) induce winding corrections:
\begin{equation}
V_{g_s}^{\rm W} \simeq -\dfrac{\kappa |W|^2}{\mathcal{V}^3}\sum_a\dfrac{C_a^w}{t^a}.
\end{equation}
We find that there are 6 such possible intersection of D7/O7 stacks and subsequently six terms in the winding-type string loop corrections \cite{Leontaris:2025hly}.

\item \textbf{Higher derivative F$^4$ corrections:} K3 divisors wrapping the D7/O7 stacks have a non-zero second Chern number $\Pi_\alpha = \int c_2 \wedge D_\alpha = 24$, leading to F$^4$ corrections \cite{Ciupke:2015msa}:
\begin{equation}
V_{{\rm F}^4} \simeq - \frac{\lambda\,\kappa^2\,|W_0|^4}{g_s^{3/2} {\cal V}^4}\, 24 \, (t^1 + t^2 + t^3).
\end{equation}

\item \textbf{Absence of certain corrections:} There are no isolated D7-brane stacks, no O3-planes (making $\overline{D3}$-brane uplifting unlikely for the current construction), and no parallel D7-brane stacks (avoiding KK-type string-loop corrections).
\end{itemize}

\noindent
Combining all contributions—including perturbative `log-loop' corrections to the K\"ahler potential, yields the complete effective potential given as below:
\begin{align}
V_{\rm eff} &\approx V_{\rm up} + \frac{\mathcal{C}_1}{\mathcal{V}^3}\left(\hat{\xi} - 4\hat{\eta} + 2\hat{\eta}\ln\mathcal{V}\right) \\
&+ \frac{\mathcal{C}_2}{\mathcal{V}^4}\left(\mathcal{C}_{w_1}\tau_1 + \mathcal{C}_{w_2}\tau_2 + \mathcal{C}_{w_3}\tau_3 + \frac{\mathcal{C}_{w_4}\tau_1\tau_2}{2(\tau_1+\tau_2)} + \frac{\mathcal{C}_{w_5}\tau_2\tau_3}{2(\tau_2+\tau_3)} + \frac{\mathcal{C}_{w_6}\tau_3\tau_1}{2(\tau_3+\tau_1)}\right) \nonumber\\
&+ \frac{\mathcal{C}_3}{\mathcal{V}^3}\left(\frac{1}{\tau_1} + \frac{1}{\tau_2} + \frac{1}{\tau_3}\right), \nonumber
\end{align}
where $\mathcal{C}_{w_\alpha}$ are complex-structure moduli dependent quantities, and $V_{\rm up}$ represents the uplifting term needed to achieve a de Sitter vacuum. In addition, the various coefficients ${\cal C}_i$'s are given by,
\bea
\label{eq:calCis}
& & \hskip-1cm {\cal C}_1 = \frac{3\,\kappa\, |W_0|^2}{4}, \quad {\cal C}_2 = \frac{4\, {\cal C}_1}{3}, \quad {\cal C}_3 = - \frac{24\, \lambda\,\kappa^2\, |W_0|^4}{g_s^{3/2}}, \quad |\lambda| =  \, {\cal O}(10^{-4}), \quad \kappa = \frac{g_s\, e^{K_{cs}}}{8\pi}. \nonumber
\eea

%%%%%%%%%%%%%%%%%%%%%%%%%%%%%%%%%%%%%%%%%%%%%%%%%%%%%%%%%%%%%%%%%%%%%%%%%%%%
\subsection{Moduli Stabilization in Perturbative LVS}
%%%%%%%%%%%%%%%%%%%%%%%%%%%%%%%%%%%%%%%%%%%%%%%%%%%%%%%%%%%%%%%%%%%%%%%%%%%%
\noindent
Considering the leading order no-scale breaking contribution in the scalar potential (\ref{eq:VpLVSsimp}) results in a non-supersymmetric AdS minimum with an exponentially large VEV for the overall volume determined by the following approximate relation \cite{Leontaris:2022rzj,Leontaris:2025hly,Leontaris:2025xit}:
\bea
\label{eq:pert-LVS}
& & \langle {\cal V} \rangle \simeq e^{\frac{13}{3}-\frac{\hat\xi}{2\, \hat\eta} -\frac{\hat\sigma}{ \hat\eta}} = e^{a/g_s^2 + b}, 
\eea
where
\bea
& & a = \frac{\zeta[3]}{2 \zeta[2]\eta_0} \simeq \frac{0.365381}{\eta_0}, \quad b = \frac{13}{3}+\frac{\sigma_0}{\eta_0}~\cdot
\eea
For natural values $\sigma_0 = -2$ and $\eta_0 = 1$, the numerical estimate for $g_s = 0.2$ gives $\langle {\cal V} \rangle = 95594.5$ while $g_s = 0.1$ leads to $\langle {\cal V} \rangle = 7.615 \cdot 10^{16}$. Further, similar to the standard Swiss-Cheese LVS case, this corresponds to a non-supersymmetric AdS minimum. Given that an exponentially large VEV of the overall volume $\cal V$ is obtained by using only the perturbative effects, this scheme is referred as ``perturbative LVS". 

After fixing the overall volume modulus ${\cal V}$ at leading order, the remaining two K\"ahler moduli can be stabilized by invoking various other subleading effects, for example, those arising from the string-loop corrections and the higher derivative F$^4$ corrections.

%%%%%%%%%%%%%%%%%%%%%%%%%%%%%%%%%%%%%%%%%%%%%%%%%%%%%%%%%%%%%%%%%%%%%%%%%%%%
%%%%%%%%%%%%%%%%%%%%%%%%%%%%%%%%%%%%%%%%%%%%%%%%%%%%%%%%%%%%%%%%%%%%%%%%%%%%

\section{Assisting Fibre Inflation with Multiple-fields}

\subsection{Single-Field Inflationary Dynamics in perturbative LVS}
\noindent
After minimizing the overall volume (${\cal V}$) of the CY threefold via the perturbative LVS scheme, one is left with two K\"ahler moduli, say $t^2$ and $t^3$, which still remain flat. In order to get an effective single-field inflationary model, one can incorporate some appropriate worldvolume gauge fluxes on suitable divisors \cite{Minasian:1997mm,Freed:1999vc}, leading to a correlation between the two moduli arising through the vanishing of FI parameter. On these lines, following the prescription of \cite{Cicoli:2017axo}, one gets $t^2 = t^3$, or equivalently $\tau_2 = \tau_3$ for some suitable gauge flux choice in our concrete model. Furthermore, considering a canonically normalized field $t^2 = e^{\phi/\sqrt3}$, the effective single-field potential $V(\phi)$ takes the following form:
\begin{equation}
\label{eq:Vsingle-field}
V(\phi) \simeq \mathcal{C}_0\left(\mathcal{R}_{\rm LVS} + \mathcal{R}_0 e^{-2\gamma\phi} - e^{-\gamma\phi} + \mathcal{R}_1 e^{\gamma\phi} + \mathcal{R}_2 e^{2\gamma\phi}\right),
\end{equation}
where we have used $\gamma=1/\sqrt{3}$ while the model dependent parameters are given as below:
\bea
\label{eq:Cis-new}
& & \hskip-0.5cm {\cal C}_0 =\frac{{\cal C}_2 \tilde{\cal C}_w}{{\cal V}^3},  \quad {\cal R}_{\rm LVS} = \frac{{\cal C}_{\rm up}}{{\cal C}_0 {\cal V}^p} +  \frac{{\cal C}_1}{{\cal C}_0 {\cal V}^3} \left(\hat\xi + 2\,\hat\eta \, \ln{\cal V} - 8\,\hat\eta + 2\,\hat\sigma \right),\\
& & \hskip-0.5cm {\cal R}_0 = \frac{{\cal C}_3}{2{\cal C}_2\tilde{\cal C}_w}, \quad \frac{{\cal R}_1}{{\cal R}_0} = \frac{4}{{\cal V}}, \quad \frac{{\cal R}_2}{{\cal R}_0} = -\frac{4\,{\cal C}_2 {\cal C}_{w_1}}{{\cal C}_3\,{\cal V}} \biggl[1-\hat{\cal C}_w \left(1+\frac{2\,e^{\sqrt{3}\phi}}{{\cal V}}\right)^{-1}\biggr],\nonumber\\
& & \hskip-0.5cm \tilde{\cal C}_w =(4{\cal C}_{w_2} + 4{\cal C}_{w_3} + {\cal C}_{w_5})/4, \quad \hat{\cal C}_w = ({\cal C}_{w_4}+{\cal C}_{w_6})/({2{\cal C}_{w_1}}). \nonumber
\eea
The uplifting parameter ${\cal C}_{\rm up}$ depends on the choice of a specific mechanism corresponding to some specific values of the parameter $p$. For example, $p$ corresponds to the uplifting parameter with $p = 4/3$ for anti-D3 uplifting \cite{Kachru:2003aw,Crino:2020qwk,Cicoli:2017axo,AbdusSalam:2022krp}, and $p = 2$ for D-term uplifting \cite{Burgess:2003ic,Achucarro:2006zf,Braun:2015pza} while  $p = 8/3$ for the T-brane uplifting \cite{Cicoli:2015ylx,Cicoli:2017shd}. Let us also note that ${\cal C}_0, {\cal R}_{\rm LVS}, {\cal R}_0$ and ${\cal R}_1$ depend on the overall volume ${\cal V}$ only, and do not have a dependence on the inflaton $\phi$ while ${\cal R}_2$ depends on both the $({\cal V}, \phi)$ moduli. However, this $\phi$ dependence is further suppressed in terms of an additional inverse volume factor, and therefore does not affect the inflationary dynamics. 

To achieve successful inflation, the potential must satisfy the slow-roll conditions, quantified by the parameters:
\begin{align}
\epsilon_V &\equiv \frac{1}{2}\left(\frac{V_\phi}{V}\right)^2 \ll 1, \\
\eta_V &\equiv \frac{V_{\phi\phi}}{V} \ll 1, \nonumber\\
\xi^{(2)}_V &\equiv \frac{V_\phi V_{\phi\phi\phi}}{V^2} \ll 1, \nonumber
\end{align}
where derivatives are with respect to the canonically normalized inflaton field $\phi$. The primary observational signatures of inflation are encoded in:
\begin{itemize}
\item The scalar spectral index $n_s$, which enters the primordial power spectrum from the scalar perturbation: $\mathcal{P}_\mathcal{R}(k) = A_s(k/k_s)^{n_s-1}$.
\item The tensor-to-scalar ratio $r$, measuring the relative amplitude of tensor perturbations.
\item The running of the spectral index $\alpha_s = dn_s/d\ln k$.
\end{itemize}
\noindent
The various cosmological observables can be directly correlated with the slow-roll parameters $\epsilon_V$ and $\eta_V$ as below \cite{Planck:2018jri},
\bea
\label{eq:cosmo-observables}
& & \hskip-1cm P_s \equiv \frac{V_{\rm inf}^\ast}{24 \pi^2 \, \epsilon_V^\ast} \simeq 2.1 \times 10^{-9}, \quad n_s - 1  \simeq 2 \, \eta_V^\ast - 6\, \epsilon_V^\ast \simeq -0.04, \quad r \simeq 16 \epsilon_V^\ast < 0.036,
\eea
where the various cosmological observables are evaluated at the horizon exit $\phi = \phi^\ast$,  such that one also has sufficient e-foldings: $N_e(\phi^\ast) \gtrsim 60$. In fact, the number of e-foldings $N_e$ depends on many things including the post-inflationary aspects as collected below, \cite{Liddle:2003as,Cicoli:2017axo}:
\bea
\label{eq:cosmo-observables1}
& & \hskip-0.5cm N_e \simeq \int_{\phi_{\rm end}}^{\phi_\ast} \frac{V}{V^\prime} d\phi \simeq 57 + \frac{1}{4} \ln(r_\ast V_\ast) - \frac{1}{3}\ln\left(\frac{10V_{\rm end}}{m_{\inf}^{3/2}}\right),
\eea
where $\phi_{\rm end}$ corresponds to end of inflation determined by $\epsilon_H = 1$ and $m_{\rm inf}$ is the inflaton mass. Also, it is worth noting that typically $N_e \simeq 50$ for Fibre inflation \cite{Cicoli:2017axo,Bhattacharya:2020gnk,Cicoli:2020bao}. A more quantified bound on the spectral index $n_s$ and its running $\alpha_s$ as suggested by the current observational constraints are:
\begin{itemize}
\item \textbf{Planck (2018)} \cite{Planck:2018jri, Planck:2018vyg} :  $n_s = 0.9651 \pm 0.0044$, \, \, $\alpha_s = -0.0041 \pm 0.0067$, \, \, $r < 0.036$.
\item \textbf{Planck+ACT+DESI} \cite{ACT:2025tim,ACT:2025fju,DESI:2024mwx, Frolovsky:2025iao} : $n_s = 0.9743 \pm 0.0034$, \, \, $\alpha_s = 0.0062 \pm 0.0052$.
\end{itemize}
\noindent
Notably, the combined Planck+ACT+DESI data disfavors negative running $\alpha_s$ and shows tension with simple Starobinsky and Higgs inflation models.

\subsubsection*{Numerical Results}
It turns out that the first three terms of the potential (\ref{eq:Vsingle-field}) correspond to the Starobinsky-like inflation \cite{Starobinsky:1980te,Brinkmann:2023eph}. Therefore, all one needs to do is to examine whether the inflationary plateau remains robust to fulfill all the cosmological requirements such as generate sufficient efolds, matching with the experimental values of the various cosmological observables, and the sub-leading corrections induced through ${\cal R}_1$ and ${\cal R}_2$ don't spoil it. In this regard, we have presented several numerical benchmark models in \cite{Leontaris:2025hly}; for example with a focus on the cosmological observabe values corresponding to the PLANCK data \cite{Planck:2018jri, Planck:2018vyg} and another one corresponding to the ACT+DESI results \cite{ACT:2025tim,ACT:2025fju,DESI:2024mwx, Frolovsky:2025iao}. For demonstration purposes, one such numerical model is given below:
\bea
\label{eq:singlefield-parameters}
& & \hskip-1cm p = 8/3, \quad \chi = -224, \quad \eta_0 = 6, \quad \sigma_0 = -4, \\
& & \hskip-1cm g_s = 0.295, \quad W_0 = 5, \quad \lambda = - 1.7 \cdot 10^{-4}, \quad {\cal C}_{\rm up} = 3.868, \nonumber\\
& & \hskip-1cm {\cal C}_{w_1} = 0.001, \quad {\cal C}_{w_2} = - 0.0008 = {\cal C}_{w_3}, \quad {\cal C}_{w_4} = -0.1 = {\cal C}_{w_6} , \quad {\cal C}_{w_5} = 0.37, \nonumber\\
& & \hskip-1cm  \nonumber\\
& & \hskip-1cm \langle {\cal V} \rangle = 1120.32, \quad \langle t^2 \rangle = 1.0267, \quad \langle \phi \rangle = 0.0456, \quad \phi_{\rm end} = 1.0945, \quad \phi^\ast = 5.75, \nonumber\\
& & \hskip-1cm N = 51.41, \quad P_s^\ast = 2.11 \cdot 10^{-9}, \quad n_s^\ast = 0.975, \quad \alpha_s^\ast = 2.75 \cdot 10^{-4}, \quad r^\ast = 3.96 \cdot 10^{-3}. \nonumber
\eea
The first block with three lines collects the various model dependent parameters, fourth line gives the moduli VEVs and the corresponding inflaton values at the horizon exit as well as at the end of inflation. The last line collects the efolds and the various cosmological observables evaluated at the horizon exit $\phi = \phi^\ast$. We observe an increase in the $n_s^\ast$ value at the cost of a slight decrease in the tensor-to-scalar ratio. Another important aspect to note is the total inflaton shift during the inflationary process, which is $\Delta\phi \simeq 5.7$M$_p$.

Further, in order to demonstrate the $\phi$ dependence of the ${\cal R}_2(\phi)$ parameter and its negligible impact on the inflationary aspects due to a volume suppression factor, we note that the effective single-field scalar potential $V(\phi)$ using the parameters in (\ref{eq:singlefield-parameters}) can be given as below
\bea
\label{eq:Vnum-single-field}
& & \hskip-1.5cm {\cal C}_0 = 2.38378\cdot10^{-10}, \qquad {\cal R}_{\rm LVS} = 0.480076, \qquad {\cal R}_0 = 0.516494,\\
& & \hskip-1.5cm {\cal R}_1 = 0.00184409, \qquad {\cal R}_2 = -0.0000196392 + \frac{2.20022}{1120.32+2e^{\sqrt{3}\phi}}, \nonumber
\eea
which clearly shows that the ${\cal R}_1$ and ${\cal R}_2$ parameters, which induce the steepening part, are sub-leading as compared to the ${\cal R}_{\rm LVS}$ and ${\cal R}_0$ parameters, and subsequently Starobinsky-like inflationary plateau remains safe.

Thus, the typical feature of the standard single-field fibre inflation can be achieved (in perturbative LVS framework) with an inflaton shift of the order of 6M$_p$. However, it has been observed in the Swiss-Cheese CY based fibre inflation models \cite{Cicoli:2018tcq,Bera:2024ihl} that such a large inflaton shift can be quite hard to achieve, if not unlikely, and this requirement usually pushes the inflaton beyond the K\"ahler cone raising the issue of overall control of the effective field theory description. This motivates one to explore the multi-field setups and engineer an assisted dynamics of multiple fields such that none of those need to be pushed too close or beyond the allowed K\"ahler cone boundaries while achieving a successful inflation phase. 

%%%%%%%%%%%%%%%%%%%%%%%%%%%%%%%%%%%%%%%%%%%%%%%%%%%%%%%%%%%%%%%%%%%%%%%%%%%%
%%%%%%%%%%%%%%%%%%%%%%%%%%%%%%%%%%%%%%%%%%%%%%%%%%%%%%%%%%%%%%%%%%%%%%%%%%%%

%%%%%%%%%%%%%%%%%%%%%%%%%%%%%%%%%%%%%%%%%%%%%%%%%%%%%%%%%%%%%%%%%%%%%%%%%%%%
%%%%%%%%%%%%%%%%%%%%%%%%%%%%%%%%%%%%%%%%%%%%%%%%%%%%%%%%%%%%%%%%%%%%%%%%%%%%

\subsection{Assisted Multi-Field Inflationary Dynamics in perturbative LVS}

%%%%%%%%%%%%%%%%%%%%%%%%%%%%%%%%%%%%%%%%%%%%%%%%%%%%%%%%%%%%%%%%%%%%%%%%%%%%
%%%%%%%%%%%%%%%%%%%%%%%%%%%%%%%%%%%%%%%%%%%%%%%%%%%%%%%%%%%%%%%%%%%%%%%%%%%%
\noindent
Taking a multi-field approach, we find that Fibre inflation naturally realizes \textit{assisted inflation}. Having multiple scalar fields collectively involved in driving the inflation helps in sharing the burden of generating sufficient efolds needed for a successful early universe cosmology, and without individual fields traveling too large super-Planckian distances in the moduli space, which may result in having some confrontational issues in the light of swampland distance conjecture. This has several advantages:
\begin{itemize}
\item Each field contributes potential energy, leading to a flatter effective potential.
\item Individual field excursions are reduced compared to single-field scenarios.
\item The multi-moduli landscape of string theory naturally provides multiple candidate inflatons and we use this possibility to exploit them to assist the inflationary scenarios
.
\end{itemize}

\noindent
The multi-field dynamics are governed by the equations of motion using the e-folding number $N$ as time coordinate ($dN = H dt$):
\begin{align}
\label{eq:FieldEquation1}
\frac{d^2\Phi^a}{dN^2} &+ \Gamma^a_{bc}\frac{d\Phi^b}{dN}\frac{d\Phi^c}{dN} + \left(3 + \frac{1}{H}\frac{dH}{dN}\right)\frac{d\Phi^a}{dN} + \frac{\mathcal{G}^{ab}\partial_b V}{H^2} = 0,
\end{align}
with the Friedmann equation:
\begin{align}
\label{eq:FieldEquation2}
H^2 = \frac{1}{3}\left(V(\Phi^a) + \frac{1}{2}H^2\mathcal{G}_{ab}\frac{d\Phi^a}{dN}\frac{d\Phi^b}{dN}\right).
\end{align}
Starting with a scalar potential given as a function of the moduli fields $\Phi^a$, the following form of the field equations (\ref{eq:FieldEquation1}) can be directly useful,
\bea
\label{eq:EOM2}
& & \frac{d^2\Phi^a}{dN^2}+{\Gamma^a}_{bc} \frac{d\Phi^b}{dN} \frac{d\Phi^c}{dN}+\left(3- \frac{1}{2} \, {\cal G}_{ab} \frac{d\Phi^a}{dN} \frac{d\Phi^b}{dN} \right) \left(\frac{d\Phi^a}{dN}+ \frac{{\cal G}^{ab} \partial_b V}{V} \right)=0~,
\eea
where the scalar potential $V$, the metric ${\cal G}_{ab}$ and ${\cal G}^{ab}$ are explicit functions of the fields $\Phi^a$. 

Subsequently, once the effective potential is determined in a given concrete global multi-field model, one can numerically solve the above second-order differential equations (\ref{eq:EOM2}), and get the evolution of the field trajectories $\Phi^a(N)$ in terms of the number of $e$-folds. Subsequently, the cosmological observables such as the scalar power spectrum $P_s$, the spectral index $n_s$, its running $\alpha_s$, and the tensor-to-scalar ratio $r$, can be expressed in terms of the $e$-folding evolution itself. These cosmological observables are defined as below
\bea
\label{eq:cosmo-observables}
& & P_s(N) = \frac{V(N)}{24\pi^2\, \epsilon(N)}, \quad n_s(N) = 1 + \frac{1}{P_s(N)} \frac{d}{dN} \,P_s(N), \\
& & \alpha_s(N) = \frac{d}{dN} n_s(N), \qquad r(N) = 16 \epsilon(N). \nonumber
\eea
All the cosmological observables are evaluated at the horizon exit $\Phi^a = \Phi^{a\ast}$ with suitable initial conditions such that one typically gets $N(\Phi^{a\ast}) \gtrsim 50$ as argued earlier. Following the constraints from the Planck 2018 data \cite{Planck:2018jri,Planck:2018vyg}, one typically needs : $P_s \simeq 2.1 \times 10^{-9}$, $n_s=0.9651\pm 0.0044$ and $\alpha_s = - 0.0041 \pm 0.0067$. In addition, the Atacama Cosmology Telescope (ACT) data gives $n_s= 0.9666 \pm  0.0077$, Planck+ACT gives $ n_s = 0.9709 \pm  0.0038$ while Planck+ACT+DESI gives $n_s = 0.9743 \pm  0.0034$ and $\alpha_s = 0.0062 \pm 0.0052$ \cite{ACT:2025tim,ACT:2025fju,DESI:2024mwx, Frolovsky:2025iao}.

\subsubsection{Evolutionary Dynamics}
\noindent
Given that overall volume modulus (${\cal V}$) is stabilized at the leading order, we will work in the real basis $\Phi^a = \{\mathcal{V}, t^2, t^3\}$ of three K\"ahler moduli. The leading order contributions to the field space metric can be subsequently computed in the new basis using the tree-level K\"ahler potential as below
\bea
K_{T_\alpha\ov{T}_\beta} (\partial_\mu T_\alpha) (\partial^\mu \ov{T}_\beta) = \frac{1}{2} {\cal G}_{ab} (\partial_\mu \Phi^a) \,(\partial^\mu\Phi^b),
\eea
which gives
\bea
\label{eq:metric-phia}
& & \hskip-1.5cm {\cal G}_{ab} = \left(\begin{array}{ccc}
\frac{1}{{\cal V}^2} &  -\frac{1}{2 t^2 {\cal V}} & -\frac{1}{2 t^3 {\cal V}} \\
& & \\
 -\frac{1}{2 t^2 {\cal V}} & \frac{1}{(t^2)^2} & \frac{1}{2 t^2 t^3} \\
& & \\
 -\frac{1}{2 t^3 {\cal V}} &  \frac{1}{2 t^2 t^3} & \frac{1}{(t^3)^2} \\
\end{array}
\right), \qquad {\cal G}^{ab} = \left(\begin{array}{ccc}
\frac{3{\cal V}^2}{2} & \frac{{\cal V}t^2}{2} & \frac{{\cal V}t^3}{2} \\
& & \\
 \frac{{\cal V}t^2}{2} & \frac{3 (t^2)^2}{2} & -\frac{t^2t^3}{2} \\
& & \\
 \frac{{\cal V}t^3}{2} &  -\frac{t^2t^3}{2}  & \frac{3 (t^3)^2}{2}\\
\end{array}
\right).
\eea
Subsequently, the non-vanishing Christoffel connections $\Gamma^a_{bc}$ are given as below,
\bea
\label{eq:Christofell}
& & \Gamma_{11}^1 = - \frac{1}{\cal V}, \qquad \Gamma_{22}^2 = - \frac{1}{t^2}, \qquad \Gamma_{33}^3 = - \frac{1}{t^3}.
\eea
Now, we consider the full scalar potential in the real field basis $\Phi^a = \{\mathcal{V}, t^2, t^3\}$ which is:
\begin{align}
V(\mathcal{V}, t^2, t^3) &= \frac{\mathcal{C}_{\rm up}}{\mathcal{V}^p} + \frac{\mathcal{C}_1}{\mathcal{V}^3}\left(\hat{\xi} + 2\hat{\eta}\ln\mathcal{V} - 8\hat{\eta} + 2\hat{\sigma}\right) \\
&- \frac{\mathcal{C}_2}{\mathcal{V}^3}\left(2\mathcal{C}_{w_1}\frac{t^2t^3}{\mathcal{V}} + \frac{\mathcal{C}_{w_2}}{t^2} + \frac{\mathcal{C}_{w_3}}{t^3} + \frac{\mathcal{C}_{w_4}t^2t^3}{\mathcal{V} + 2(t^2)^2t^3} + \frac{\mathcal{C}_{w_5}}{2(t^2+t^3)} + \frac{\mathcal{C}_{w_6}t^2t^3}{\mathcal{V} + 2t^2(t^3)^2}\right) \nonumber\\
&+ \frac{\mathcal{C}_3}{\mathcal{V}^3}\left(\frac{1}{2t^2t^3} + \frac{t^2}{\mathcal{V}} + \frac{t^3}{\mathcal{V}}\right) + \cdots \nonumber
\end{align}
This potential exhibits exchange symmetry $t^2 \leftrightarrow t^3$ and depends on the uplifting mechanism characterized by the $p$ parameter such that $p=2$ corresponds to D-term uplifting, $p=8/3$ corresponds to T-brane uplifting, and $p=4/3$ corresponds to $\overline{D3}$-brane uplifting.

%Having the model specific pieces of information at hand, one can explicitly write the second order differential equations governing the evolutionary dynamics.
Using the  model-specific data, we can now explicitly write the second-order differential equations that describe the evolutionary dynamics.
 These are given as follows,
\bea
\label{eq:Explicit-EOMs}
& & \hskip-1cm {\cal V}^{\prime\prime} = \frac{{\cal V}^{\prime2}}{{\cal V}} - \left(3- \epsilon\right) \left({\cal V}^{\prime}+\frac{3{\cal V}^2}{2V} \partial_{\cal V} V +\frac{{\cal V} \, t^2}{2V}\partial_{t^2} V +\frac{{\cal V} \, t^3}{2V}\partial_{t^3} V \right),\\
& & \hskip-1cm (t^2)^{\prime\prime} = \frac{(t^2)^{\prime2}}{t^2} - \left(3- \epsilon\right) \left((t^2)^{\prime}+\frac{{\cal V} \, t^2}{2V}\partial_{{\cal V}} V +\frac{3(t^2)^2}{2V} \partial_{t^2} V -\frac{t^2 t^3}{2V}\partial_{t^3} V\right),\nonumber\\
& & \hskip-1cm (t^3)^{\prime\prime} = \frac{(t^3)^{\prime2}}{t^3} - \left(3- \epsilon\right) \left((t^3)^{\prime}+\frac{{\cal V} \, t^3}{2V}\partial_{{\cal V}} V -\frac{t^2 t^3}{2V}\partial_{t^2} V +\frac{3(t^3)^2}{2V} \partial_{t^3} V \right).\nonumber
\eea
Here as earlier, the prime $^\prime$ denotes derivatives with respect to the number of $e$-folds $N$, i.e. ${\cal V}^\prime = \frac{d{\cal V}}{dN}$ etc, and now the inflationary parameter $\epsilon$ takes the following explicit form,
\bea
& & \hskip-1cm \epsilon = \frac{1}{2} \left(\frac{{\cal V}^{\prime2}}{{\cal V}^2} + \frac{(t^2)^{\prime2}}{(t^2)^2}  + \frac{(t^3)^{\prime2}}{(t^3)^2} - \frac{{\cal V}^\prime \, (t^2)^{\prime}}{{\cal V} \, t^2} - \frac{{\cal V}^\prime \, (t^3)^{\prime}}{{\cal V} \, t^3} + \frac{(t^2)^\prime \, (t^3)^{\prime}}{t^2\, t^3} \right).
\eea

\subsubsection{Numerical Analysis Validating the Assisted Nature of Inflation}

\noindent
In order to determine the evolutionary trajectories of various fields, we now solve the second-order differential equation~(\ref{eq:Explicit-EOMs}) under the following initial conditions:
\bea
& & \Phi^a(0)=\Phi^a_0 \qquad  {\rm and} \qquad \frac{d\Phi^a}{dN}|_{N=0}=0\,.\eea
After solving the evolution equations numerically, we have presented several benchmark models compatible with observational data \cite{Leontaris:2025hly} with and/or without incorporating the recent ACT+DESI results. For demonstration purposes, here we present one of those models:
\bea
\label{eq:model-M3-3-ACT}
& & \hskip-1.5cm p = 8/3, \qquad \chi({\rm CY}) = -224, \qquad \eta_0 = 6, \qquad \sigma_0 = -4, \qquad g_s = 0.295, \\
& & \hskip-1.5cm  |W_0| = 5, \qquad {\cal C}_{w_1} = 0.001, \qquad  {\cal C}_{w_2} = -0.0008, \qquad  {\cal C}_{w_3} = -0.0008,    \nonumber\\
& & \hskip-1.5cm {\cal C}_{w_4} = -0.1, \qquad {\cal C}_{w_5} = 0.33, \qquad  {\cal C}_{w_6} = -0.1, \qquad \lambda = - 0.00017; \nonumber\\
& & \nonumber\\
& & \hskip-1.5cm {\cal C}_{\rm up} = 5.32455, %\quad \langle V \rangle =  9.26 \cdot 10^{-23}, 
\quad \langle {\cal V} \rangle = 1123.23, \quad \langle t^2 \rangle = 1.14996, \quad \langle t^3 \rangle = 1.14996, \nonumber\\
& & \hskip-1.5cm \langle \phi^1 \rangle = 6.01802, \quad \langle \phi^2 \rangle = -2.41341  , \quad \langle \phi^3 \rangle = -2.41341 , \nonumber\\
& & \nonumber\\
& & \hskip-1.5cm {\cal V}^\ast = 1258.22, \qquad (t^2)^\ast = 25.8, \quad  (t^3)^\ast = 25.8, \nonumber\\
& & \hskip-1.5cm \phi^{1\ast} = 6.11069, \quad \phi^{2\ast} = 1.35 , \quad \phi^{3\ast} = 1.35, \quad \Delta\phi = 5.32, \quad N = 55.5, \quad N^\ast = 5.5,\nonumber\\
& & \hskip-1.5cm  P_s^\ast =  2.095\cdot 10^{-9}, \quad n_s^\ast = 0.9763, \quad \alpha_s^\ast = -5.763\cdot10^{-4}, \quad r^\ast = 2.73\cdot10^{-3}.\nonumber
\eea
The effective inflaton shift $\Delta\phi$ can be considered as the distance in the flat three-dimensional field space spanned by canonical fields ${\phi^a}$, and it turns out that
\bea
\label{eq:shiftphi-three-field}
& & \Delta \phi = \sqrt{\left(\phi^{1\ast} - \langle\phi^1\rangle\right)^2 + \left(\phi^{2\ast} - \langle\phi^2\rangle\right)^2 + \left(\phi^{3\ast} - \langle\phi^3\rangle\right)^2}~,
\eea
which is simply the Pythagorean distance between two points. Here, $\phi^{a\ast}$ corresponds to the horizon exit while $\langle \phi^a\rangle$ denotes the moduli VEVs at the perturbative LVS minimum. In fact, for any choice of canonical field basis, the inflaton shift $\Delta\phi$ remains invariant. However, the individual fields may result in different shifts in a different choice of basis. Since our scalar potential has exchange symmetry $2 \leftrightarrow 3$ inherited from the CY threefold, we have used the following basis of canonical fields $\phi^a$
\bea
\label{eq:cononical-varphi3}
& & \phi^1 = \frac{1}{\sqrt{3}} \left(\varphi^1+ \varphi^2 + \varphi^3 \right) = \sqrt{\frac{2}{3}} \ln(\sqrt{2}\,{\cal V}) , \\
& & \phi^2 = \frac{1}{6} \left(2\sqrt{3}\, \varphi^1 + (3 -\sqrt{3})\varphi^2-(3+\sqrt{3} \phi^3) \right), \nonumber\\
& & \phi^3 = \frac{1}{6} \left(2\sqrt{3}\, \varphi^1 - (3 +\sqrt{3})\varphi^2 + (3 - \sqrt{3}) \phi^3 \right),\nonumber
\eea
where 
\bea
\label{eq:cononical-varphi0}
& & \varphi^a = \frac{1}{\sqrt{2}} \ln \tau_a, \qquad \forall \, a \in \{1, 2, 3\}.
\eea
For the benchmark model presented in (\ref{eq:model-M3-3-ACT}), we have
\begin{equation}
\Delta\phi^1 \simeq 0.0926, \qquad \Delta\phi^2 = \Delta\phi^3 \simeq 3.763 \qquad \Rightarrow \qquad \Delta\phi \simeq 5.32.
\end{equation}
This illustrates a key advantage of assisted inflation: while single-field models require $\Delta\phi \simeq 6 M_p$, the two-field approach reduces individual excursions to be around $3.5 M_p$, mitigating concerns about trans-Planckian displacements.

\begin{figure}[H]
\centering
\includegraphics[width=15.1cm]{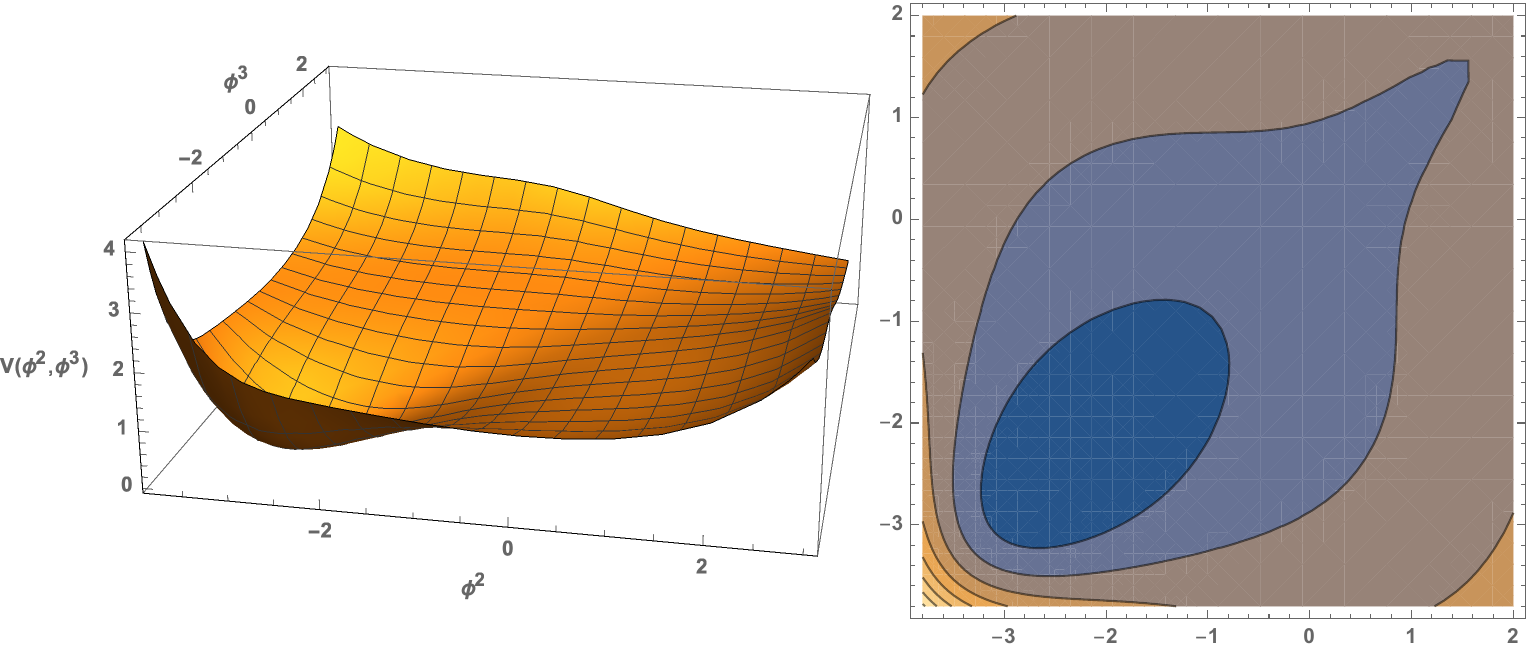}
\caption{Assisted inflationary track in the $(\phi^2, \phi^3)$ plane while keeping $\phi^1$ fixed at its minimum \cite{Leontaris:2025hly}.}
\label{fig_pot-3D-three-field}
\end{figure}

\noindent
Assisted nature of the inflationary track can be seen from figure \ref{fig_pot-3D-three-field}. For fixed volume $\mathcal{V}$, the potential $V(\phi^2, \phi^3)$ exhibits a flat direction along the diagonal, ideal for slow-roll inflation. The relation $\Delta\phi^n \simeq \Delta\phi/\sqrt{n}$ for $n=2$ fields demonstrates how assisted inflation naturally addresses the eta-problem and large-field concerns.

%%%%%%%%%%%%%%%%%%%%%%%%%%%%%%%%%%%%%%%%%%%%%%%%%%%%%%%%%%
\subsubsection{Robustness: Stability and Scale Separation}
The inflationary dynamics is governed by a variety of scalar potential pieces arising from different sources at different orders in the perturbation series. For example, it turns out that the individual scalar potential contributions at the LVS minimum are qualitified as below,
\bea
\label{eq:various-V-vevs1}
& & \hskip-1cm \langle V_{\rm up} \rangle = 2.83449 \cdot 10^{-8}, \quad \langle V_{\rm logloop} \rangle = -3.48662\cdot 10^{-8}, \quad \langle V_{\alpha^\prime} \rangle = 6.61081 \cdot 10^{-9}, \\
& & \hskip-1cm \langle V_{g_s}^W \rangle = -1.82453 \cdot 10^{-10}, \quad \langle V_{{\rm F}^4} \rangle = 9.28818\cdot 10^{-11}, \nonumber
\eea
which implies that there is no clean hierarchy among the various individual contributions! However, we note that the hierarchy is maintained among pieces driving inflation and those which do not get involved in the inflationary purpose, i.e., $|V_{\rm inf}| \ll |V_{\rm pLVS}|$ where one may define $V_{\rm inf} = V_{g_s}^W + V_{{\rm F}^4}$ and $V_{\rm pLVS} = V_{\alpha^\prime} + V_{\rm logloop}$. In addition, the following eigenvalues of the mass-squared matrix indeed ensure a mass hierarchy among the various moduli: $m_a^2 = \{4.76\cdot 10^{-9}, \, 1.91\cdot 10^{-10}, \, 6.27 \cdot 10^{-11}\}$. The heaviest eigenstate corresponds to the overall volume modulus ${\cal V}$ while the other ones are some combinations of all the three moduli $\{{\cal V}, t^2, t^3\}$. 

\begin{figure}[H]
\centering
\includegraphics[width=15.5cm]{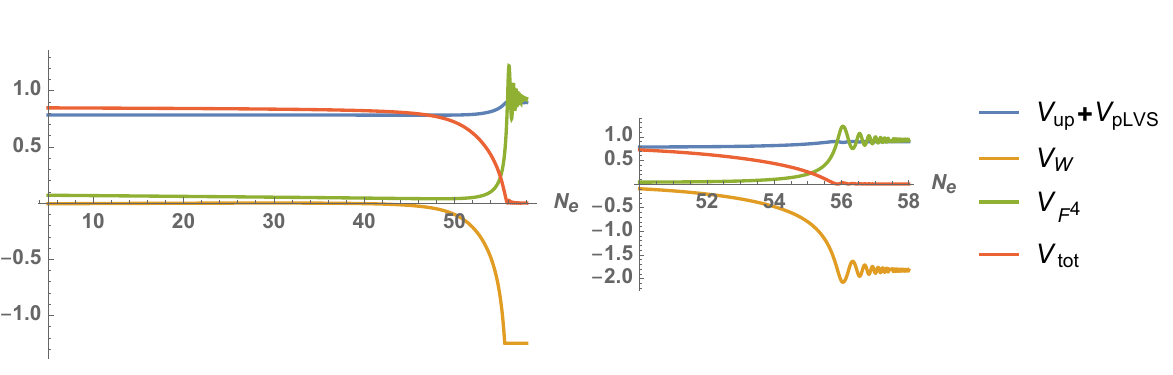}
\caption{Evolution of various pieces of the scalar potential ($V\cdot 10^{10}$) contributions \cite{Leontaris:2025hly}.}
\label{fig_V-scalesc-2-three-field}
\end{figure}

\noindent
During the entire evolutionary process, the various mass scales must respect the mass hierarchy : $H < m_{3/2} < M_{\rm KK} < M_s < M_p$, where $H$ is the Hubble scale and the gravitino mass is denoted as $m_{3/2}$. The string mass $M_s$ and the various KK scales are defined as below 
\bea
\label{eq:KKscales}
& & \hskip-1cm M_s = \frac{M_p}{\sqrt{\alpha^\prime}}, \quad M_{\rm KK}^a = \frac{M_p}{R_a} = \frac{M_s}{\tilde{R}_a}, \quad m_{3/2} = e^{\frac{1}{2} {\cal K}} |W_0| = \frac{\sqrt{g_s}\, |W_0|}{\sqrt{2} \, {\cal V}},
\eea
where using $\ell_s = 2\pi \sqrt{\alpha^\prime}$ results in $M_s = \frac{g_s^{1/4} \sqrt\pi}{\sqrt{\cal V}} M_p$ \cite{Conlon:2006gv}, and $R_a = \tilde{R}_a \sqrt{\alpha^\prime}$ where $\tilde{R}_a$ is the size of the relevant length corresponding to a particular KK mode such that $\tilde{R}_a = (t^a)^{1/2}$ for two-cycle volumes, $\tilde{R}_a = (\tau_a)^{1/4}$ for four-cycle volumes and $\tilde{R}_L = {\cal V}^{1/6}$  corresponds to the bulk modulus ($t^b \simeq{\cal V}^{1/3}$ or $\tau_b \simeq {\cal V}^{2/3}$). We denote the bulk KK mode mass as $M_{\rm KK}^b$ while $M_{\rm KK}^1$ corresponds to $t^1$ modulus and $M_{\rm KK}^{2,3}$ correspond to $t^2$ and $t^3$ moduli. 

\begin{figure}[H]
\centering
\includegraphics[width=15.6cm]{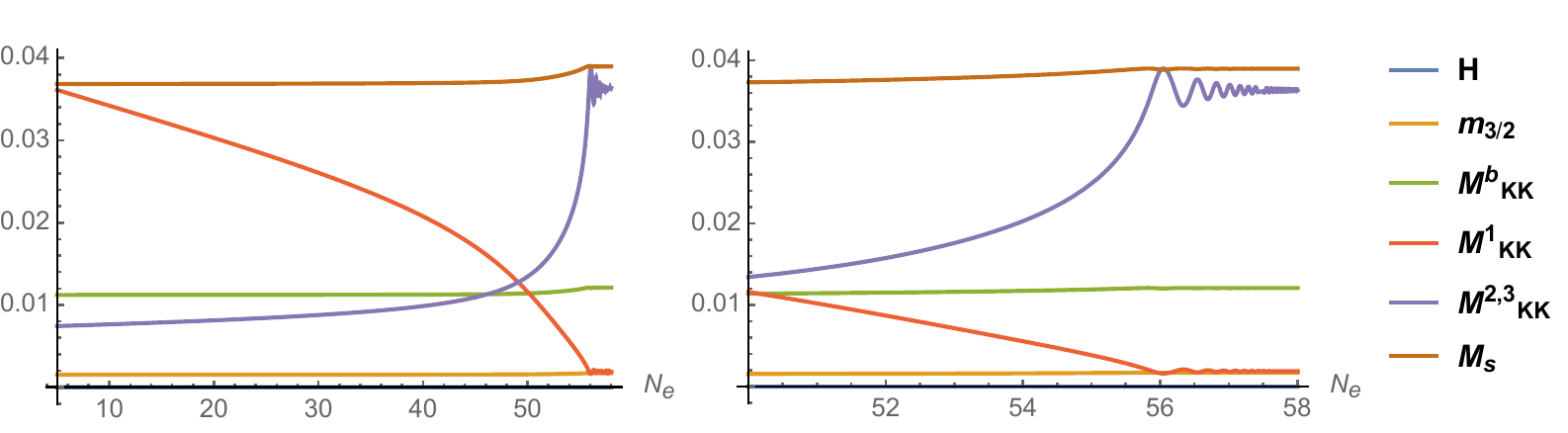}
\caption{Evolutions of various mass scales during inflationary dynamics \cite{Leontaris:2025hly}}
\label{fig_scales-three-field}
\end{figure}

\noindent
The evolution of various scales as presented in figure \ref{fig_scales-three-field} shows that the desired mass-hierarchy is respected throughout the inflationary regime, i.e. till $\epsilon \leq 1$ corresponding to $N\simeq55.5$. However, we also observe that the heaviest KK scales $M_{\rm KK}^{2,3}$ become comparable to the string mass towards the minimum, after the end of inflation. Moreover, towards the horizon exit when the inflatons $t^2$ and $t^3$ are shifted significantly away from their minimum, the $t^1$ modulus tends towards smaller size and subsequently, the corresponding KK scale, namely $M_{\rm KK}^1$, increases and becomes closer to the string mass scale $M_s$ towards the horizon exit $N^\ast = 5.5$. However, as long as the volume moduli do not enter in the regime where they take too smaller values, the supergravity approximations should remain (marginally) valid. It is also worth noting that the string mass scale being comparable to (one of the) KK mass may not be an immediate concern as argued in \cite{Dienes:2002ze}.

%%%%%%%%%%%%%%%%%%%%%%%%%%%%%%%%%%%%%%%%%%%%%%%%%%%%%%%%%%%%%%%%%%%%%%%%%
%%%%%%%%%%%%%%%%%%%%%%%%%%%%%%%%%%%%%%%%%%%%%%%%%%%%%%%%%%%%%%%%%%%%%%%%%

\section{Conclusions }

In this work we  presented a successful global embedding of assisted fibre inflation within the perturbative Large Volume Scenario (pLVS) in type IIB string theory. It is first shown that in the pLVS framework K\"ahler  moduli are stabilized without relying on non-perturbative effects, making it applicable to a wider class of Calabi-Yau manifolds. We have presented a novel realization of multi-field fibre inflation (assisted inflation) which allows sufficient efolds without requiring any single modulus to exceed its K\"ahler cone bound. We have achieved  embedding  the model  in a simple K3-fibred Calabi-Yau compactification manifold with Hodge numbers $h^{1,1}=3$, $h^{2,1}=115$. We  stabilized the volume at $\mathcal{V} \simeq 10^3$ with $g_s \simeq 0.3$ and $W_0 \simeq 5$ and introduced D-terms (or T-branes) to implement the uplifting mechanism, this way ensuring a de Sitter vacuum. For reasonable values of our parameters, such as the string coupling and fluxes, we find that our  predictions are compatible with Planck/ACT/DESI cosmological data. We further point out an interesting fact that the assisted two-field inflation reduces the individual field excursions to $\simeq 3.5 M_p$ as compared to the single-field requirement of around $6 M_p$, presenting a way to address the trans-Planckian concerns, e.g. those raised in \cite{Cicoli:2018tcq,Bera:2024ihl}, by increasing the number of moduli in the game. 

In summary, we reviewed the recent work \cite{Leontaris:2025hly} on multi-field fibre inflation in perturbative LVS where we discussed that the burden of creating sufficient efolds can be shared by multiple fibre moduli. This result can be of significant interest in the light of recent challenges faced in standard single-field fibre inflation models of standard Swiss-Cheese LVS \cite{Cicoli:2018tcq,Bera:2024ihl}. Global embedding of such multi-field scenario in perturbative LVS demonstrates that fiber inflation represents a robust, globally consistent framework for inflation in string theory, naturally addressing moduli stabilization while producing observationally viable predictions. 

%%%%%%%%%%%%%%%%%%%%%%%%%%%%%%%%%%%%%%%%%%%%%%%%%%%%%%%%%%%%%%%%%%%%%%%%%%%%%%%%%%%%%%%%%%
%%%%%%%%%%%%%%%%%%%%%%%%%%%%%%%%%%%%%%%%%%%%%%%%%%%%%%%%%%%%%%%%%%%%%%%%%%%%%%%%%%%%%%%%%%

\subsection*{Acknowledgments}
\noindent
We are thankful to all our collaborators for the earlier collaborations on the related subject, in particular, in the area of moduli stabilization and inflationary (global) model building. PS is thankful to the {\it Department of Science and Technology (DST), India} for the kind support.

%%%%%%%%%%%%%%%%%%%%%%%%%%%%%%%%%%%%%%%%%%%%%%%%%%%%%%%%%%%%%%%%%%%%%%%%%%%%%%%%%%%%%%%%%%
%%%%%%%%%%%%%%%%%%%%%%%%%%%%%%%%%%%%%%%%%%%%%%%%%%%%%%%%%%%%%%%%%%%%%%%%%%%%%%%%%%%%%%%%%%

%\newpage
%\bibliographystyle{JHEP}
%\bibliography{reference}

\newpage
\bibliographystyle{utphys}
\bibliography{reference}

\end{document}